\documentclass[letterpaper,twocolumn,10pt]{article}
\usepackage{usenix,epsfig,endnotes}
\usepackage{amsmath, xcolor}
\begin{document}

\date{}

\title{\Large \bf Training Through Failure: Effects of Data Consistency in Parallel Machine Learning Training}

\newcommand{\equalcontrib}{\textsuperscript{1}}
\newcommand{\equalcontribnote}{\endnote{ Equal Contribution.}}

\newcommand{\modelinfo}{\textsuperscript{2}}
\newcommand{\modelinfonote}{\endnote{ The CNN model used in this research consists of two convolutional layers (with 16 and 32 filters, respectively), each followed by a ReLU activation and a max pooling layer. After the convolutional layers, the feature maps are flattened and fed into a fully connected layer with 512 units, followed by another ReLU activation. A dropout layer with a 25\% dropout rate is then applied to prevent overfitting. Finally, the output layer consists of 10 units, representing the 10 classes in the FashionMNIST dataset.}}

\author{
{\rm Ray Cao\equalcontrib}\\
Stanford, Google
\and
{\rm Sherry Luo\equalcontrib}\\
Stanford, Google
\and
{\rm Steve Gan\equalcontrib}\\
Stanford, Google
\and
{\rm Sujeeth Jinesh\equalcontrib}\\
Stanford, Google
} 

\maketitle

\thispagestyle{empty}

\subsection*{Abstract}
In this study, we explore the impact of relaxing data consistency in parallel machine learning training during a failure using various parameter server configurations. Our failure recovery strategies include traditional checkpointing, chain replication (which ensures a backup server takes over in case of failure), and a novel stateless parameter server approach. In the stateless approach, workers continue generating gradient updates even if the parameter server is down, applying these updates once the server is back online. We compare these techniques to a standard checkpointing approach, where the training job is resumed from the latest checkpoint.

To assess the resilience and performance of each configuration, we intentionally killed the parameter server during training for each experiment. Our experiment results indicate that the stateless parameter server approach continues to train towards convergence and improves accuracy as much as 10\% in the face of a failure despite using stale weights and gradients. The chain replication and checkpointing techniques demonstrate convergence but suffer from setbacks in accuracy due to restarting from old checkpoints. These results suggest that allowing workers to continue generating updates during server downtime and applying these updates later can effectively improve hardware utilization. Furthermore, despite higher resource usage, the stateless parameter server method incurs similar monetary costs in terms of hardware usage compared to standard checkpointing methods due to the pricing structure of common cloud providers.

\section{Introduction}

Modern Machine Learning training on the scale of ChatGPT \cite{openai2022language}, Claude \cite{anthropic2023claude}, Gemini \cite{google2023introducing}, etc. often use specialized hardware accelerators (e.g. GPUs, TPUs). Models are often trained at "MegaScale" \cite{jiang2024megas}, exceeding 10,000 hardware accelerators. Due to the scale of training and the preemptive environment of cloud computing \cite{yang2023skypilot}, long running training jobs often encounter failures.

In the event of a failure, training jobs often need to be restarted from a checkpoint potentially up to hours old, incurring significant startup overhead. Moreover, some existing approaches leave accelerators idle while awaiting the recovery of failed peers, leading to resource wastage and increased costs.

An additional result of unplanned failures is that computation, and therefore progress towards a convergent model stops. This is because of researchers desire to have consistent and repeatable model updates even though this level of consistency may not be necessary.

This paper aims to address these issues by introducing techniques designed to optimize hardware utilization and progress towards convergence in environments prone to preemptions and failures, particularly focusing on data parallelism in distributed training. 

There two common ways to parallelize model training: model parallelism and data parallelism. Model parallelism partitions a single model into disjoint subsets of parameters and assigns each subset of parameters to one dedicated training instance \cite{dean2012large}. Data parallelism splits up a large dataset into many batches and sends them across multiple workers to generate gradients in parallel for weight updates at one or multiple parameter servers \cite{goyal2017accurate} \cite{krizhevsky2012imagenet} \cite{sergeev2018horovod}.

In synchronized data parallel training, workers pull from the same snapshot of weights from the parameter server and push gradients back to the parameter server after processing data. The parameter server then collects and applies these gradients collectively. It has been observed that model training can still continue even if we don't always fetch the latest weights\cite{li2014scaling}. 

Building on this observation, we explore the potential of relaxing consistency requirements for both workers receiving weights and the parameter server receiving gradients. Our goal is to achieve convergence progress from parameter server failures while simultaneously optimizing hardware utilization during these disruptions.

\section{Design}

\subsection{Sync/Async Checkpointing}

We use synchronous and asynchronous data parallel training with checkpointing for failure recovery, both featuring a centralized parameter server. The parameter server saves a checkpoint of parameters to persistent storage after a set number of weight updates.

Weight updates differ between synchronous and asynchronous data parallel training:

\begin{itemize}
\item \textbf{Synchronous Data Parallel Training}: Enforces strict synchronization before each iteration, aggregating gradients from all workers before updating weights. This ensures consistency across iterations but can be inefficient due to slow workers and their associated tail latency.
\item \textbf{Asynchronous Data Parallel Training}: Allows workers to compute and submit gradients independently, with the parameter server updating weights as gradients become available. This improves parallelism and resource usage but may suffer from stale weights in workers. This is due to potential network differences between workers, or heterogenous workers where some workers are significantly more powerful than others in the same cluster.
\end{itemize}

When a parameter server dies, the remaining workers are idle for both experiments, but when the parameter server is resurrected, it will look for the latest checkpoint in its store and rehydrate the model's weights and continue from that older state.
\begin{figure}[h]
\begin{center}
\includegraphics[width=1.0\linewidth]{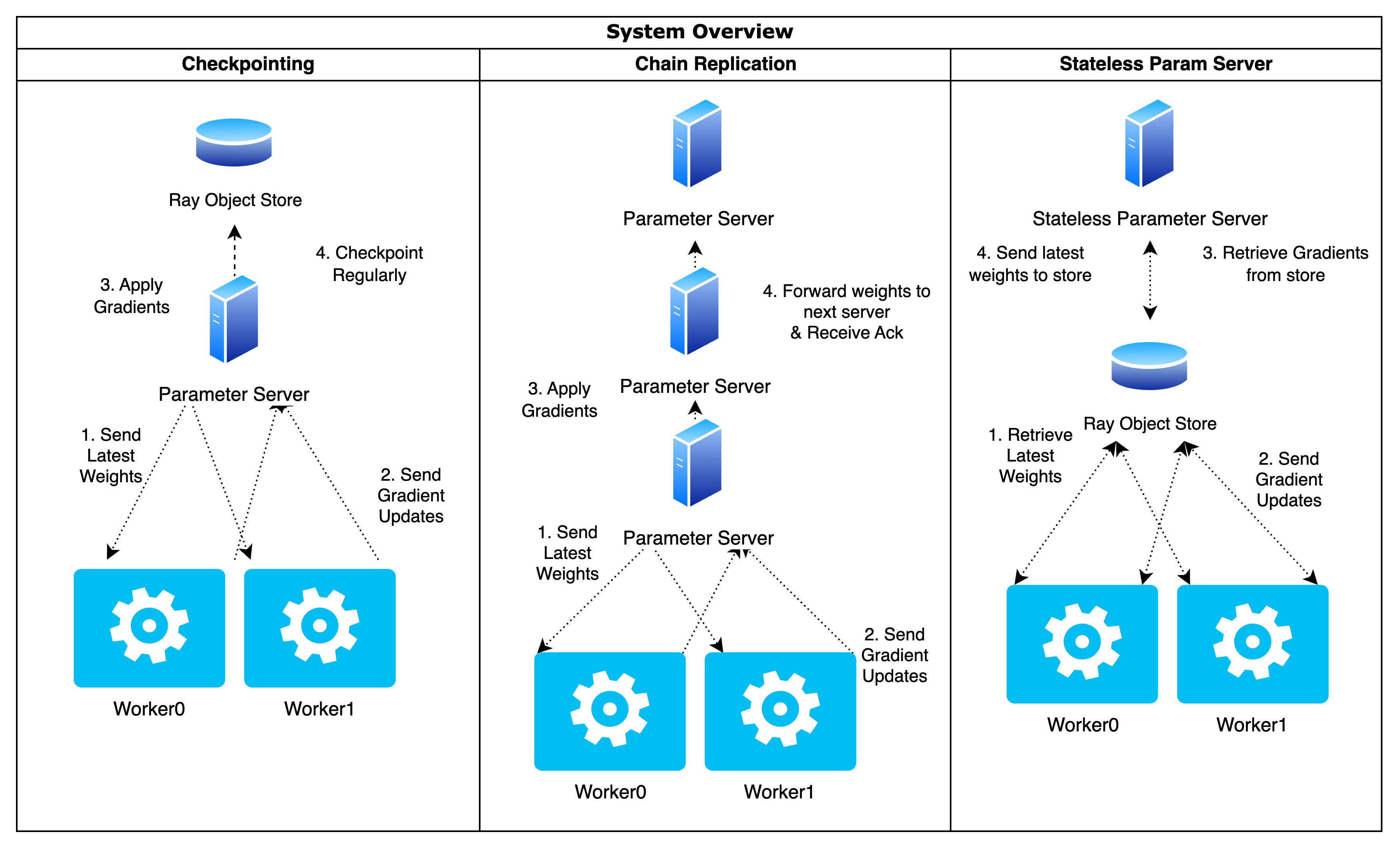}
\end{center}
\caption{System Overview}
\end{figure}
\subsection{Sync/Async Chain Replicated Parameter Servers}

In this experiment, we use a chain of servers to replicate model parameters and achieve fast failure recovery. There are two major differences between our design and  traditional chain replication \cite{van2004chain}, specifically in:
\begin{itemize}
    \item \textbf{Relaxed Consistency}: Traditional chain replication provides a strong data consistency guarantee by requiring the replica chain to forward a data update from the head (frontend) to the tail before considering the update as successful. In contrast, our design only waits for acknowledgment from the next server in the chain during replication.
    \item \textbf{Periodic Replication}: Unlike traditional replication, which occurs with every client update, we replicate weights only after a predetermined number of weight updates to improve speed during regular training.
\end{itemize}

We relaxed the consistency requirement because of the data transfer overhead involved in replicating model weights across servers, which can be a slow process. Moreover, we don't anticipate simultaneous failures of multiple servers back-to-back. In the rare case that both the frontend parameter server and the secondary parameter fail, the tertiary parameter server will recover the training process with more outdated weights.

Compared to checkpointing, this approach incurs the overhead of maintaining multiple parameter servers. However, it offers the advantage of immediate training resumption upon failure, as the new frontend server already has the weights warm  in memory.

\begin{figure}[h]
\begin{center}
\includegraphics[width=1.0\linewidth]{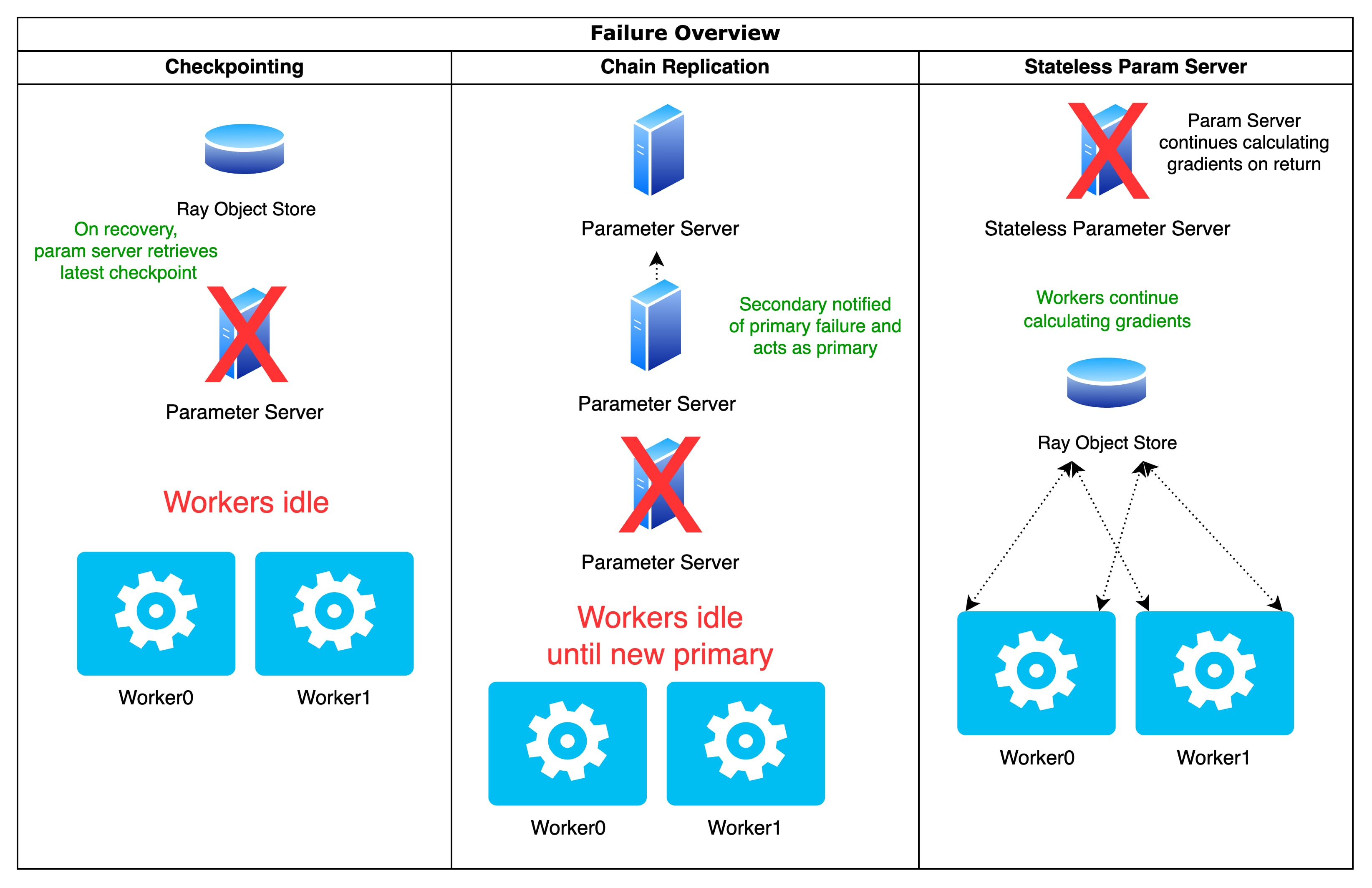}
\end{center}
\caption{Failures Overview}
\end{figure}

\subsection{Stateless Parameter Server}

The traditional parameter server is stateful because it stores weights updated in every training iteration. This means the weights share the same fate as their hosting server—if the parameter server crashes unexpectedly, the weights are lost. By making the parameter server stateless—a function that retrieves weights, applies gradient updates, and then stores the new weights—we can achieve failure recovery through task re-execution.

To make the parameter server stateless, we offload the weights to an external persistent store, removing them from the server. This allows the parameter server to function as a stateless task, rather than a stateful actor. By using a distributed in-memory store \cite{roy2020memory}, we can achieve memory-level access speeds while maintaining the fault tolerance of a distributed file store, ensuring efficient weight sharing with workers.

For the workers, we relax the data consistency requirement, allowing them to continue training on the latest weights that they have until the parameter server recovers. If the parameter server crashes, workers will retrieve a weight snapshot from persistent storage and continue pushing gradients computed on new data based on this snapshot. When the parameter server is resurrected, it will apply the interim, stale gradients. This approach maintains high hardware utilization for workers that would otherwise be idle.

Experimentally, we found that tuning the learning rate down for a large number of pending gradients facilitated training progress. Additionally, exploring alternative methods for applying gradients for the stale gradients accumulated during the parameter server downtime could lead to further accuracy improvement. For example, gradient clipping \cite{zhang2019gradient}, elastic averaging SGD \cite{zhang2015deep} and adaptive learning rate \cite{zeiler2012adadelta}.

\section{Implementation}

We implement this project in Python3 and we use Ray \cite{moritz2018ray} to implement parameter servers and workers for all experiments. We implement the coordination for chain replication and the asynchronous stateless parameter server experiments using the Kazoo library for Zookeeper \cite{zookeeper}.

\subsection{Sync/Async Checkpointing}

We implement the synchronous and asynchronous checkpointing approach using Ray's actor paradigm. A centralized $ParameterServer$, a Ray actor, orchestrates the training process by maintaining the model's weights and aggregating gradients calculated by multiple remote worker tasks of $compute\_gradients$. Performance is evaluated periodically; accuracy and loss metrics collected by a $MetricExporter$ actor.

We simulate failures in all experiments by killing the actor using Ray's $kill$ function that SIGTERMs the respective experiment's process id.

In the synchronous checkpointing experiment, the parameter server actor spawns training workers as Ray tasks in each iteration. It waits until all worker gradients are ready and then applies all updates to the model parameters at once, thus ensuring consistent model states across iterations. This achieves the strongest data consistency level of all the experiments.

In the asynchronous checkpointing experiment, workers operate independently, continuously fetching model weights from the parameter server and computing gradient updates. The parameter server updates the model parameters as soon as it receives any worker gradient. This asynchronous technique is slightly less data consistent than the synchronous parameter server.

\subsection{Sync/Async Chain Replication}
We implement the parameter server as a Ray actor maintaining weights as its states with stateless worker tasks as in Sync/Async Checkpointing. 

We use Zookeeper to facilitate communications among the servers. Each parameter server $i$ maintains a Zookeeper client that creates an ephemeral znode $z_i$ and keeps a watch on the ephemeral znode $z_{i - 1}$ of the previous parameter server. If it detects a failure on the previous parameter server via the Zookeeper watch, it either discovers a new server $j$ with $j < i - 1$ as its previous server or establishes itself as the new frontend server. 

All ephemeral znodes created by parameter servers' Zookeeper clients are child nodes of a base node $z_b.$ The Zookeeper client learns about the global node states by querying for the child nodes of the base node.

After a predetermined number of weight updates, the frontend parameter server puts its current weights into the Ray's object store and sends the \textit{reference} to the weights as the data to its znode. The secondary parameter server in line receives a notification via the Zookeeper watch for the data update, fetches the weights from the object store, and then sets its own weights. It then sends the same object reference as the data to its own znode to further propagate weights down the chain.

During a failure, the server chain encounters an idle period until the new frontend parameter server has been established. 

\subsection{Async Stateless Parameter Server}

For the asynchronous stateless parameter server, we also implement it as a Ray actor but purely for gathering metrics. We keep it stateless otherwise to emulate a stateless task.

Weights are gradients are both stored in Ray's object store. We create two Zookeeper znodes $/weights$ and $/gradient\_updates$ to store references to weights and gradients, respectively.

In a single stateless parameter server step, we gather all references to gradients from the $/gradient\_updates$ znode and obtain the weights using the reference in the $/weights$ znode. After applying gradients to the weights, the task updates the $/weights$ znode with the reference to the new weights.

In a worker step, we retrieve the weights from the $/weights$ znode and generate gradients based on this snapshot of weights with new data. Subsequently, the worker adds the reference to the new gradients to the $/gradient\_updates$ znode.

\begin{figure}[!ht]
{\tt \small
\begin{verbatim}
@ray.remote()
def stateless_parameter_server():
    gradient_updates = zk.get_all("/gradient_updates")
    weights = zk.get("/weights")
    latest_weights = apply_gradients(weights,
        gradient_updates)
    zk.put("/weights", latest_weights)
\end{verbatim}
}

{\tt \small
\begin{verbatim}
@ray.remote()
def stateless_worker():
    latest_weights = zk.get("/weights")
    gradient_update = trainer.apply(latest_weights)
    zk.append("/gradient_updates", gradient_update)
\end{verbatim}
}
\caption{Pseudo-code describing the Stateless Parameter Server experiment}
\end{figure}

Gradient update accesses to the store can theoretically be lock-free, as the updated gradients are written only once by the worker, and the update is read and later deleted by the parameter server once it's done applying the update. However, Kazoo's implementation of Zookeeper doesn't allow this, so we implement a simple zlock to prevent race conditions and missed gradient updates. The locking mechanism forces a worker or parameter server to acquire the lock before putting in gradients, or updating the weights. Accesses to weights can also be made with a reader-writer lock, but was not done in this implementation.

\section{Evaluation}

In the evaluation section, the results presented were obtained by training a convolutional neural network \cite{o2015introduction} (CNN) model\modelinfo \ on the FashionMNIST \cite{xiao2017fashion} dataset. All experiments were conducted in a testing environment with an M1 Max 10-core CPU (8 performance, and 2 efficiency), 64GB unified memory, and 2-TB SSD storage \cite{apple2020specs}. 

\begin{figure}[!ht]
\includegraphics[width=1.0\linewidth]{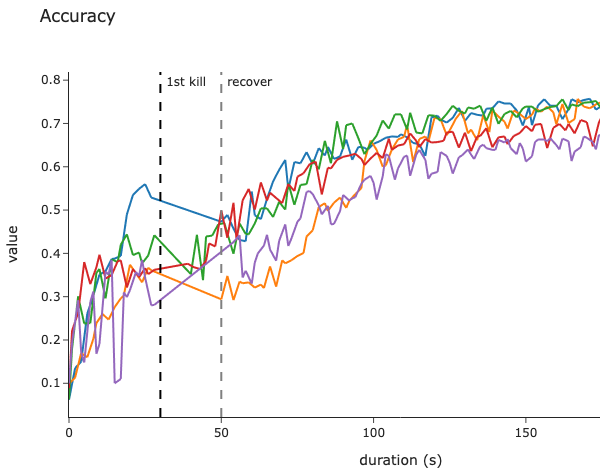}
\caption{Training accuracy after killing and recovering once. \\  Legend: \textcolor{blue}{Blue} - Sync checkpointing, \textcolor{orange}{Orange} - Async checkpointing, \textcolor{green}{Green} - Sync chain replication, \textcolor{red}{Red} - Async chain replication, \textcolor{violet}{Purple} - Stateless parameter server}
\label{accuracy_1}
\end{figure}

In Figure \ref{accuracy_1}, we see that the accuracy of chain replication, along with stateless parameter server remain relatively constant even after one kill. Our higher fault-tolerant methods do not need long periods of recovery time and steadily increase in accuracy. Meanwhile, we observe a noticeable drop in accuracy for the checkpointing experiments.

Although synchronous checkpointing initially exhibits a faster increase in accuracy compared to the other experimental methods, it subsequently suffers a large drop in accuracy due to the interruption from failure. The initial increase in accuracy may be attributed to the data consistency resulted from applying gradients in order. We note that both synchronous and asynchronous checkpointing methods demonstrate a lack of resilience to training failures, as evidenced by the drop in accuracy.

\begin{figure}[!ht]
\includegraphics[width=1.0\linewidth]{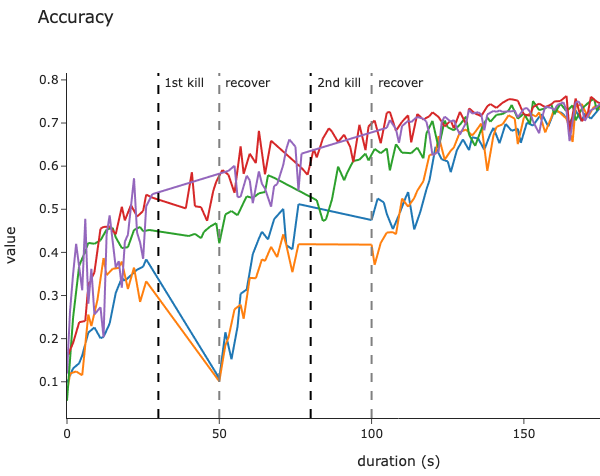}
\caption{Training accuracy after killing and recovering twice.  Legend: \textcolor{blue}{Blue} - Sync checkpointing, \textcolor{orange}{Orange} - Async checkpointing, \textcolor{green}{Green} - Sync chain replication, \textcolor{red}{Red} - Async chain replication, \textcolor{violet}{Purple} - Stateless parameter server}
\label{accuracy_2}
\end{figure}

We observe the greater fault-tolerance capabilities of our newer experiments---chain replication and stateless parameter server---especially after a series of two kill-and-recover sequences. In \ref{accuracy_2}, we notice that checkpointing struggles to recover in time before the second kill. Meanwhile, our fault-tolerant strategies continue training without major disruptions from parameter server crashes.

Over the course of many runs, we also noticed that stateless parameter server consistently achieved higher accuracies after the parameter server recovers than when compared to before it died, sometimes as high as a 15\% leap in accuracy, indicating that stale gradients can still contribute significantly towards model convergence. This is likely due to the various gradient updates that all, when applied, would push the weights closer to the local minima.

As the number of parameter server failure increases, we notice that the performance gap in failure recovery between the checkpoining methods and the chain replication and stateless server methods becomes more apparent. Following a single failure, checkpointing demonstrates prompt recovery. However, with the occurrence of a second failure, the checkpointing approach lags significantly behind the more fault-tolerant strategies. It takes an additional 10\% of time to attain a comparable level of accuracy achieved by the chain replication and stateless parameter server methods, reaching approximately 70\% accuracy.
 
\subsection{Costs Discussion}

While our accuracy figures demonstrate that chain replication and stateless parameter server show higher fault tolerance than checkpointing, we compare resource utilization of three failure recovery mechanisms in terms of cloud computing costs.

Typical cloud contracts involve long-term agreements of holding hardware accelerators for a fixed time and cost, so it's in the user's best interest to keep the hardware running at as close to maximum utilization for as long as possible \cite{google2024tpupricing}.

\begin{figure}[!ht]
\includegraphics[width=1.0\linewidth]{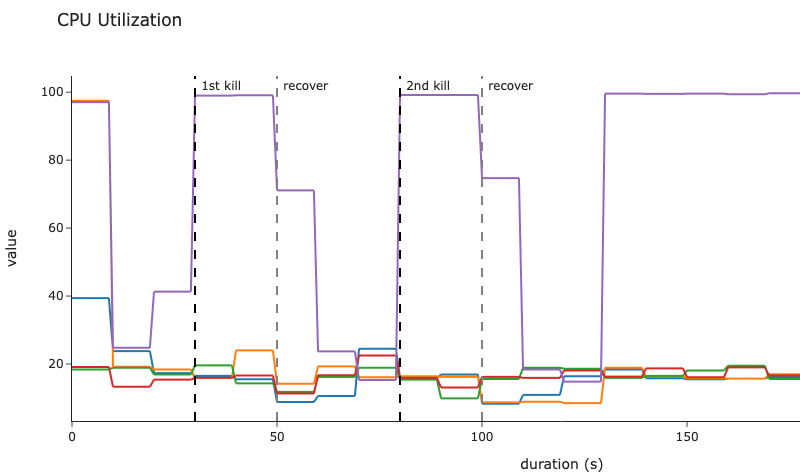}
\caption{CPU utilization after killing and recovering twice. \\ Legend: \textcolor{blue}{Blue} - Sync checkpointing, \textcolor{orange}{Orange} - Async checkpointing, \textcolor{green}{Green} - Sync chain replication, \textcolor{red}{Red} - Async chain replication, \textcolor{violet}{Purple} - Stateless parameter server}
\label{cpu_utilization}
\end{figure}

\begin{figure}[!ht]
\includegraphics[width=1.0\linewidth]{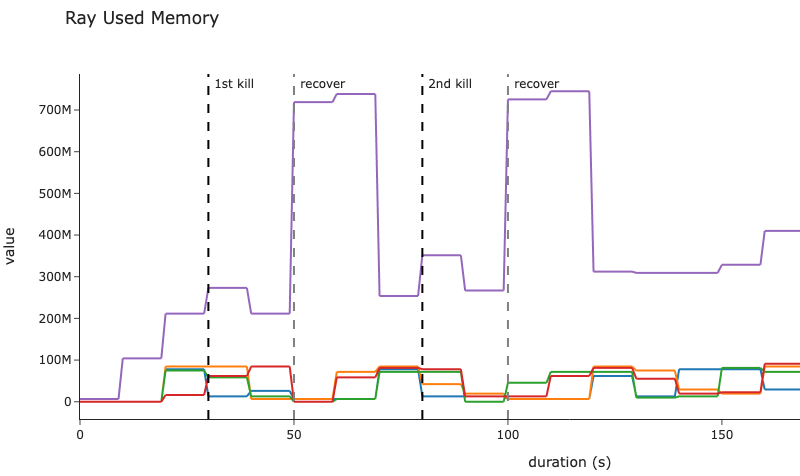}
\caption{Memory utilization after killing and recovering twice. \\ Legend: \textcolor{blue}{Blue} - Sync checkpointing, \textcolor{orange}{Orange} - Async checkpointing, \textcolor{green}{Green} - Sync chain replication, \textcolor{red}{Red} - Async chain replication, \textcolor{violet}{Purple} - Stateless parameter server}
\label{memory}
\end{figure}

\begin{figure}[!ht]
\includegraphics[width=1.0\linewidth]{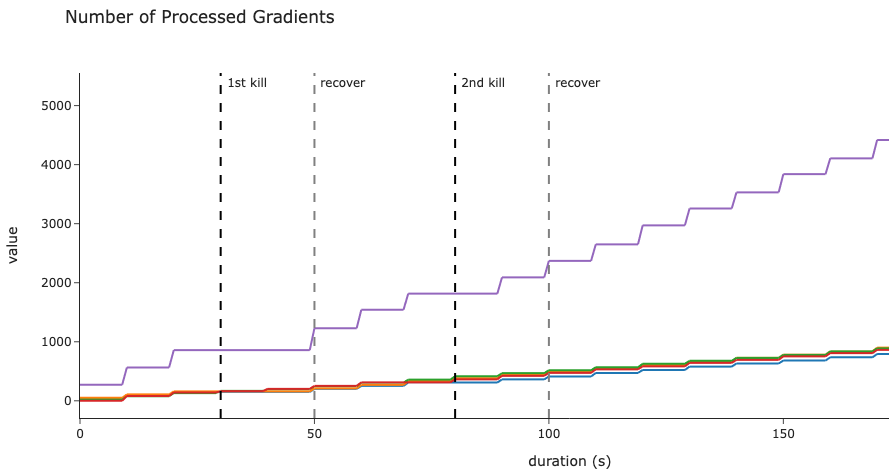}
\caption{Number of gradients processed after killing and recovering twice. \\ Legend: \textcolor{blue}{Blue} - Sync checkpointing, \textcolor{orange}{Orange} - Async checkpointing, \textcolor{green}{Green} - Sync chain replication, \textcolor{red}{Red} - Async chain replication, \textcolor{violet}{Purple} - Stateless parameter server}
\label{processed_gradiants}
\end{figure}

For chain replication, we see that for effectively every resource usage metric (Figures \ref{cpu_utilization}, \ref{memory}, and \ref{processed_gradiants}), the experiments are comparable (as expected) to checkpointing. 

For stateless parameter server, we see a clear trend in extra hardware utilization. In Figure~\ref{cpu_utilization} (CPU utilization), because the parameter server is not present at the very beginning of the experiment and during the crash periods, there is no synchronization costs between the parameter server and the workers. The workers simply compute gradients and perform asynchronous puts for gradients during this period. We achieve almost full hardware utilization even during the parameter server's downtime. This workload becomes CPU-bound rather than I/O-bound with the parameter server. When the parameter server recovers, it notifies the workers and performs a synchronization step. Because calls to get weights are synchronous, workers slow down training. This results in a sharp decrease in the CPU utilization.

Compared to our checkpointing and chain replication designs where each training iteration involves spawning new workers, the stateless parameter server approach uses the same workers for the entirety of the experiment. Because of the less overhead of spawning workers, we can generate more gradients with this approach as shown in Figure~\ref{processed_gradiants}. Note that because we implement this metric using Ray's custom metric interface, which is asynchronous, the number of iterations may not align exactly with the value.

We also note that the stateless parameter server approach has a higher memory usage compared to other approaches. During the server recovery time, the parameter server reads and apply all gradients generated by the workers during its downtime which generates spikes in Ray's Used Memory metric as shown by Figure~\ref{memory}.

These results show that the stateless parameter server excels at maintaining high hardware utilization during server failures. Asynchronous workers can use stale weights to continue training, and our approach leverages stale gradients to sustain progress during downtime. Despite high resource usage, it matches the costs of checkpointing and chain replication under the fixed pricing structure. While it doesn't surpass other methods in accuracy, it demonstrates a promising direction for fault-tolerant training by enabling continued training through failures with stale gradients.

\section{Related Work}

Researchers have developed various techniques and mechanisms \cite{verbraeken2020survey} to help distributed machine learning systems handle failures, including checkpointing \cite{chen2014numarck}, replication \cite{lee2017speeding}, and lineage recovery \cite{zhang2017diagnosing}. Frameworks like Ray \cite{moritz2018ray} offer fault tolerance through lineage-based recovery and task reconstruction, whose native mechanism primarily focuses on task-level failures instead of parameter server failures. Additionally, other traditional methods like checkpointing have limitations in terms of lost progress and accuracy setbacks. Our research explores alternative approaches, including server chain replication and a stateless parameter server to maximally preserve training progress.

Asynchronous training \cite{amiri2020machine}, including the notable Hogwild! algorithm \cite{recht2011hogwild}, has gained its popularity in achieving high throughput in parallel machine learning scenarios. Nevertheless, its reliance on strict consistency and default fault tolerance provided by underlying infrastructure (i.e. MapReduce \cite{dean2008mapreduce}) can pose challenges to model convergence in the face of failures. Our approach aims to bridge this gap by merging the advantages of asynchronous updates with enhanced robustness to failures. This allows workers to persistently generate updates even when the major parameter server is unavailable, leading to improvements in training efficiency and hardware utilization. 

\section{Future Work}
There are several action items for future work to develop a comprehensive framework for resilient and efficient parallel machine learning training with our research:

\begin{itemize}
    \item \textbf{Applying the system to diverse machine learning models and tweaking model parameters}: Extending this research to natural language processing and reinforcement learning applications would increase the generalizability of our research.  There are some modeling concerns with stateless parameter server. This technique may not hold well in a situation without normalization layers, or could result in heavy loss of accuracy for larger models.

    \item \textbf{A more in-depth analysis of techniques in a cloud environment}: Understanding how network conditions influence the performance of our research would be crucial for its real-world deployment in distributed environments. Testing the compatibility between our stateless parameter server and hardware accelerators, such as GPUs and TPUs, would lead to further performance gains and cost reductions.

    \item \textbf{More efficient implementation of stateless parameter server}: As noted in the evaluation section, chain replication demonstrated excellent fault tolerance and resilience to failures by keeping weights warm in the backup servers' memory. Meanwhile, the stateless parameter server maintained high hardware utilization and kept accelerators active. We can combine these two approaches and utilize the backup servers as storage for weights and gradients. Also improving the locking mechanism in the stateless parameter server approach would help with speed.
\end{itemize}

\section{Conclusion}

This study demonstrates that relaxing data consistency in parallel machine learning model training can enhance fault tolerance and efficiency. We evaluated checkpointing, chain replication, and a stateless parameter server approach, each offering distinct advantages.

The stateless parameter server approach, which allows workers to continue generating and applying gradient updates during server downtime, showed significant promise. It maintained training momentum during failure while incurring similar costs to traditional checkpointing methods due to efficient hardware utilization.

Overall, enabling continuous worker updates during server failures helps maintain training progress and reduce costs. Future research should explore these techniques across various models, analyze their network impact, and integrate them with hardware accelerators to further optimize distributed training.

\newpage
\bibliography{main}
\theendnotes
\equalcontribnote
\modelinfonote
\end{document}